\documentclass[journal]{IEEEtran}
\IEEEoverridecommandlockouts

\usepackage{graphicx, subcaption}
\usepackage{amsmath,amsthm,amsfonts,amssymb}
\usepackage{cite}
\usepackage{bm}
\usepackage{bbm}
\usepackage{url}
\usepackage{array}
\usepackage{color}
\usepackage{multirow}
\usepackage{booktabs,multirow}
\usepackage{xcolor}
\usepackage{enumerate}
\usepackage{enumitem}
\usepackage{makecell}
\usepackage{comment}

\usepackage[english]{babel}
\usepackage{algorithm}
\usepackage[noend]{algpseudocode}

\theoremstyle{plain}

\usepackage{mathtools}

\usepackage[english]{babel}
\usepackage{algorithm}
\usepackage[noend]{algpseudocode}

\allowdisplaybreaks[4]

\begin{document}
\title{Over-the-Air Inference over Multi-hop MIMO Networks}

\author{
Chenghong~Bian,~\IEEEmembership{Student Member,~IEEE},
Meng~Hua,~\IEEEmembership{Member,~IEEE}, and 
Deniz~G\"und\"uz,~\IEEEmembership{Fellow,~IEEE}
\thanks{The authors are with the Department of Electrical and Electronic Engineering, Imperial College London, London SW7 2AZ, U.K. (e-mail: \{c.bian22,m.hua,d.gunduz\}@imperial.ac.uk).
}
\thanks{
This work received funding from the UKRI for the projects AI-R (ERC Consolidator Grant, EP/X030806/1) and the SNS JU project 6G-GOALS under the EU’s Horizon program Grant Agreement No. 101139232.}
}

\maketitle

\begin{abstract}
A novel over-the-air machine learning framework over multi-hop multiple-input and multiple-output (MIMO) networks is proposed. The core idea is to imitate fully connected (FC) neural network layers using multiple MIMO channels by carefully designing the precoding matrices at the transmitting nodes.
A neural network dubbed \textit{PrototypeNet} is employed consisting of multiple FC layers, with the number of neurons of each layer equal to the number of antennas of the corresponding terminal.
To achieve satisfactory performance, we train PrototypeNet based on a customized loss function consisting of classification error and the power of latent vectors to satisfy transmit power constraints, with noise injection during training.
Precoding matrices for each hop are then obtained by solving an optimization problem. We also propose a \textit{multiple-block} extension when the number of antennas is limited.
Numerical results verify that the proposed over-the-air transmission scheme can achieve satisfactory classification accuracy under a power constraint. The results also show that higher classification accuracy can be achieved with an increasing number of hops at a modest signal-to-noise ratio (SNR).
\end{abstract}


\section{Introduction}\label{sec:intro}
There has been a growing trend toward the integration of computation and wireless communications to enhance the performance of future Internet of Things (IoT) networks\cite{edge_comp_over, image_retrieval}. 
Instead of traditional digital schemes where the computation and communication are separated, over-the-air computation (OAC)  enables integrated communication and computation using analog transmissions \cite{oac_fed, oac_amiri, oac_iot, oac_mimo}. A key advantage of the OAC scheme is to shift the computation burden to the wireless domain by leveraging the propagation principle of the wireless signals. For instance, linear operations such as summations and polynomials can be carried out by exploring the signal-superposition property of the wireless medium, which have been widely applied in federated learning \cite{oac_fed, oac_amiri}. 

Recently, edge learning \cite{edge_comp_over} has been widely studied, where the nodes participate in a machine learning task and communicate with each other via wireless links. 
%The precoder was designed based on maximizing the communication utility, e.g., minimizing the computation minimum mean square of the (MSE) \textbf{\color{red}(provide references)}, however, the ultimate goal of the task is to maximize the machine learning utility, e.g., maximizing the classification accuracy \textbf{\color{red}(provide references)}. To reach the ultimate machine learning goal
To facilitate this, the concept of over-the-air neural networks (NNs) is proposed, where the output of the wireless channel is treated as the NN output \cite{air_ris,air_fc, air_mimo}. 
% By deploying reconfigurable intelligent surfaces (RIS) and designing precoders, the operations for both convolutional neural networks (CNNs) and FC layers can be realized. 
For example, the authors in \cite{air_ris} constructed a convolutional neural network for a modulation classification problem by leveraging the ambient wireless propagation environment through a reconfigurable intelligent surface.
Inspired by the similarity in the input-output relation between a multiple-input and multiple-output (MIMO) channel with a precoding matrix and a fully connected (FC) layer, the authors in \cite{air_fc} and \cite{air_mimo} developed OAC schemes for efficient deep learning algorithms over wireless networks. To be specific, in \cite{air_fc}, an OAC FC layer was constructed under the scenario where multiple synchronized users precode their messages and deliver them to the destination node, which forms a multiple input single output (MISO) system. System-level validation was conducted and a $92\%$ classification accuracy was achieved when evaluated on the MNIST dataset. However, its performance is limited by the single antenna at the receiver.
A more complicated OAC FC layer based on the MIMO channel was considered in \cite{air_mimo} for split learning, where the FC layer is constructed using the precoding and combining matrices at the transceiver.  The gradients were back-propagated via the MIMO channel to update the precoding matrix leading to extensive computational overhead, which is not suitable for low-complexity edge devices.  Moreover, these works focus on single-hop scenarios.
% Moreover, \cite{air_mimo} required updating the system parameters for different MIMO channel implementations leading to additional overheads.

\begin{figure*}[t]
  \centering
  \includegraphics[width=0.95\linewidth]{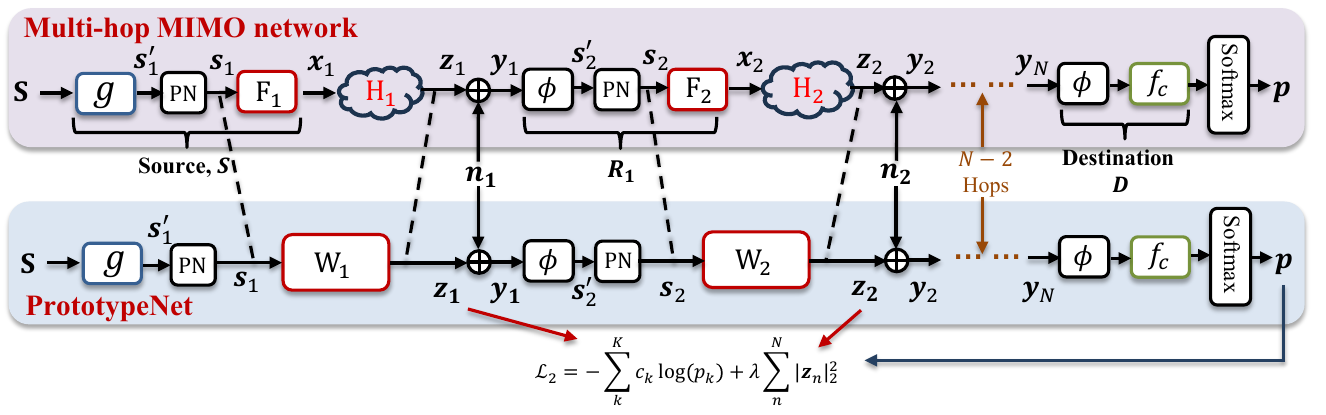}\\
  \caption{The flowchart of the multi-hop MIMO network and the PrototypeNet. We consider a $N$-hop MIMO network and it is equivalent to the PrototypeNet with $(N+1)$ FC layers (including the one at the destination node). The second loss function, $\mathcal{L}_2$, defined in \eqref{eq:optim_z} is used to train the PrototypeNet.}
\label{fig:system_model}
\end{figure*}

To fill this gap, this paper studies OAC machine learning over multi-hop MIMO networks.  A multi-hop MIMO network resembles a complex-valued neural network with multiple FC layers.  We first design a novel PrototypeNet consisting of multiple FC layers, power normalization, and non-linear activation functions, which is trained in an end-to-end fashion using a customized loss function, a combination of the classification error and the power of the latent vectors.
After obtaining the weights of PrototypeNet as well as the channel state information (CSI), the precoding matrix for each hop is obtained by solving an optimization problem, where the goal is to approximate the PrototypeNet operations over-the-air as accurately as possible. 
This differs from \cite{air_mimo}, where the precoding and combining matrices are trained along with the neural network weights.  
Numerical experiments show that the proposed OAC scheme can achieve satisfactory classification accuracy under a power constraint. The results also show that a larger number of hops leads to a higher classification accuracy even at a modest SNR value.
While this method is successful in deploying NNs over-the-air, the width of the NN is limited by the number of antennas at the transmitters. To overcome this limitations, we propose transmission of multiple channel symbols per input sample over time.
In the proposed scheme, a large-dimensional input is partitioned into several low-dimensional components, each of which is precoded by a different precoding matrix.
%We further consider a practical scenario where the nodes have a limited number of antennas compared to the input/output dimension of the FC layer in PrototypeNet. To approximate PrototypeNet with a limited number of antennas, we propose transmission of multiple channel symbols over time per input symbol. 
In this paper, we show that adopting multiple symbol transmissions can significantly improve the classification performance by allowing the deployment of a wider NN using a limited number of antennas. %This can be seen as similar to coding gain in conventional communication systems.

\section{System Model}\label{sec:II}
We consider an image classification problem over a multi-hop MIMO network, but the results can be applied to other tasks carried out by NNs. An image, denoted by ${\bf{S}} \in \mathbb{R}^{C\times H \times W}$, where $C$, $H$ and $W$ represent the number of channels, the height, and the width of the image, respectively is available at the source node ($\mathrm{S}$). Each image $\mathbf{S}$ belongs to a certain class, which can be represented via a one-hot vector $\bm{c} \in \mathbb{R}^K$, where $K$ is the total number of classes. 
Destination node ($\mathrm{D}$) would like to detect the class of the image. However, $\mathrm{D}$ does not have access to the image, and is connected to $\mathrm{S}$ only through $(N-1)$ relay nodes denoted by  $\{\mathrm{R}_1, \ldots, \mathrm{R}_{N-1}\}$.
The source, destination, and relay nodes are all equipped with multiple antennas. We further assume that the CSI of each hop is available at the transmitter. The flowchart of the OAC multi-hop MIMO network is shown in  Fig. \ref{fig:system_model} and is detailed as follows.

The source node transforms the input image $\mathbf{S}$ into a complex-valued vector denoted by $\bm{s}_1^\prime \in \mathbb{C}^{M_1}$  via a function $g(\cdot)$ which can be realized by 2D average pooling operation, where $M_1$ denotes the number of antennas at the source node. It is worth mentioning that we denote the number of antennas at the $i$-th relay node and the destination node by $M_{i+1}, i\in[1, N-1]$  and $M_{N+1}$, respectively.  Then, we apply power normalization (PN) to the complex vector $\bm{s}_1^\prime$ to obtain $\bm{s}_1$, expressed as:
\begin{equation}
    \bm{s}_1 = (\bm{s}_1^\prime - \mu_1)/v_1,
    \label{eq:s1}
\end{equation}
where $\mu_1$ and $v_1$ denote the  mean and variance of $\bm{s}_1^\prime$, respectively,  which are recorded during training. The PN module ensures that $\mathbb{E}\left[||\bm{s}_1||^2_2\right] \le M_1$. The source node $\mathrm{S}$ adopts a precoding matrix $\mathbf{F}_1 \in \mathbb{C}^{M_1 \times M_1}$, and the transmitted signal, denoted by $\bm{x}_1$, can be expressed as:
\begin{equation}
    \bm{x}_1 = \mathbf{F}_1 \bm{s}_1.
\end{equation}

After passing the channel, the received signal $\bm{y}_1 \in \mathbb{C}^{M_2}$ at the first relay node $\mathrm{R}_1$ can be expressed as:
\begin{align}
    \bm{y}_1 = \mathbf{H}_1 \bm{x}_1 + \bm{n}_1,
    \label{eq:mimo_channel}
\end{align}
where $\mathbf{H}_1$ stands for the channel between Source and Relay 1, and each element of which is independent and identically distributed (i.i.d.) following a complex Gaussian distribution, i.e., $\mathcal{CN}(0, 1)$, while $\bm{n}_1$ denotes the noise received at Relay 1 and satisfies $\bm{n}_1 \sim \mathcal{CN}(\bm{0}, \sigma^2_1 \mathbf{I}_{M_2})$. 

At $\mathrm{R}_1$, the received signal $\bm{y}_1$ is first fed to the non-linear activation function  $\phi(\cdot)$, which is implemented as a complex ReLU function. To be precise, both the real and imaginary parts of $\bm{y}_1$ are passed through standard ReLU functions. Power normalization is applied to the non-linear activation output to generate $\bm{s}_2$. We emphasize that the PN layer is essential to meet the power constraint of the relay node, $\mathrm{R}_1$.
Then the transmitted signal of the first relay node can be expressed as: 
\begin{align}
    {\bm{x}_2} = {\mathbf{F}_2}  {\rm PN}(\phi(\bm{y}_1)),
    \label{eq:operation_hops}
\end{align}
where $\mathbf{F}_2$ is the precoding matrix at $\mathrm{R}_1$ and ${\rm PN}$ denotes power normalization function. We note that the remaining relay nodes follow the same procedure and the transmit signal, the CSI and the channel noise at the $i$-th hop satisfy $\mathbb{E}\left[||\bm{s}_i||^2_2\right] \le M_i$, $\bm{H}_i \in \mathbb{C}^{M_{i+1}\times M_i}$ and $\bm{n}_i \sim \mathcal{CN}(\bm{0}, \sigma^2_i \mathbf{I}_{M_{i+1}})$, respectively.

The received signal ${\bm{y}}_{N}$ at the destination is first fed into the complex ReLU function $\phi(\cdot)$ and is transformed into a real-valued $2M_{N+1}$-dimensional vector by concatenating the real and imaginary parts of the complex vector. After passing the FC layer $f_c(\cdot)$ followed by softmax activation, a probability vector $\bm{p} \in \mathbb{R}^K$ is obtained, and the index, $k^*$, of the largest element in $\bm{p}$ is the final classification output. The classification accuracy of the system is defined as:
\begin{align}
    {\rm Acc} (\%) = \mathbb{E}(\bm{1}(k^* = c^*)),
    \label{eq:accuracy}
\end{align}
where $\bm{1}(\cdot)$ is the indicator function and $c^*$ denotes the correct class index.
In the following, we assume that the number of antennas and the noise power are identical over the hops, i.e., $M_1 = \cdots = M_{N+1} \triangleq M$ and $\sigma_1^2 = \cdots = \sigma_{N+1}^2 \triangleq \sigma^2$, for simplicity. 

\section{Methodology}\label{sec:III}
In this section, we first introduce the motivation and the training methodology of the proposed solution. Then, the procedure to obtain the precoding matrices $\{\mathbf{F}_1, \cdots, \mathbf{F}_{N}\}$ will be presented.

\subsection{PrototypeNet}\label{sec:IIIA}
In the proposed solution, we want the $i$-th hop of the relay network to act as the $i$-th layer of a target neural network. As shown in  Fig. \ref{fig:system_model}, for the $i$-th hop with MIMO channel realization, $\mathbf{H}_i$, we aim to determine a precoding matrix $\mathbf{F}_i$ such that the received signal at the $i$-th relay node can be treated as the output of a FC layer (without the bias term). 
Let $\mathbf{W}_i \in \mathbb{C}^{M_{i+1}\times M_i}$ represent the weight matrix of the $i$-th layer of the target neural network called the PrototypeNet.
The received signal of the $i$-th hop can be expressed as:
\begin{align}
    \bm{y}_{i} &= \mathbf{H}_i\mathbf{F}_i \bm{s}_i + \bm{n}_i.
    \label{eq:Fi_vs_Wi}
\end{align}
Since we want the relay network to behave like the PrototypeNet, the problem can be formulated as finding $\mathbf{F}_i$ such that $\mathbf{H}_i\mathbf{F}_i$ is equivalent to $\mathbf{W}_i$. We will first introduce a novel end-to-end optimization algorithm to determine the weights of PrototypeNet, $\mathbf{W}_i$, as detailed next.

To start with, we construct the PrototypeNet, illustrated in Fig. \ref{fig:system_model}, which resembles the multi-hop MIMO network. In particular, PrototypeNet and the multi-hop MIMO network share the same transformation function $g(\cdot)$, the PN modules, and the final FC layer, $f_c(\cdot)$.
At first glance, the PrototypeNet is identical to a standard image classification network. However, the weights obtained by training a standard image classification network fail to achieve satisfactory performance when deployed over the multi-hop MIMO network. This is because the standard network does not take the Frobenius norms of $\mathbf{W}_i, i \in [1, N]$, into account. Moreover, channel noise is not considered when training a  standard network, which degrades its performance when implemented over a noisy channel.
%In particular, each element of $\bm{W}_i$ can be scaled by an arbitrary factor without influencing the final reconstruction performance\footnote{This is because the PN module is capable to offset the effect of the scaling operations.}.  Though it is possible for the standard image classification network to control the power of the latents $\bm{z}_i$ by scaling the weights, $\bm{W}_i$, it would lead to sub-optimal performance as the latents may not be robust to noise. 
%Moreover, depending on different initialization, the weights, $\bm{W}_i$ obtained by the standard classification network can be quite different in terms of the frobneous norms. This leads to problems when deployed in the real wireless network due to the randomness of the weights, $\bm{W}_i$. 

The proposed PrototypeNet takes the power constraint as well as the channel noise into account. For the $i$-th node, after generating the complex vector $\bm{s}_i$ as in \eqref{eq:s1}, the output $\bm{y}_i$ for the next node is obtained as:
\begin{align}
    \bm{y}_{i} = \bm{z}_i + \bm{n}_i,
    \label{eq:yi}
\end{align}
where we define $\bm{z}_i \triangleq \mathbf{W}_i \bm{s}_i$.
Intuitively, there exists a trade-off between the power consumption of $\bm{z}_i$  and the classification accuracy: when $\|\bm{z}_i\|_2^2$ is large, the noise $\bm{n}_i$ is negligible and has little impact on the accuracy. On the other hand, when $\|\bm{z}_i\|_2^2$ is small, the noise dominates and the performance degrades significantly.

To achieve  power and classification accuracy trade-offs, we consider two loss functions to train PrototypeNet:
\begin{align}
    \mathcal{L}_1 &= -\sum_{k=1}^K c_k \log(p_k) +  \frac{\lambda}{N} \sum_{j=1}^N \|\mathbf{W}_j\|_F^2, \label{eq:optim_W} \\
   \text{and} \quad \mathcal{L}_2 &= -\sum_{k=1}^K c_k \log(p_k) +  \frac{\lambda}{N} \sum_{j=1}^N \|\bm{z}_j\|_2^2, \label{eq:optim_z}
\end{align}
where $\lambda$ balances the power and the classification accuracy for both loss functions. Note that the second term of  $\mathcal{L}_1$, i.e., $\frac{\lambda}{N} \sum_{j=1}^N \|\mathbf{W}_j\|_F^2$, minimizes the Frobenius norm of the weights which is analogous to the widely used weight regularization terms to avoid over-fitting. In our case, this term implicitly minimizes the power of $\bm{z}_i$. The second loss function, $\mathcal{L}_2$, on the other hand, directly minimizes $\|\bm{z}_i\|_2^2$. As we will show though simulations training the PrototypeNet using $\mathcal{L}_2$ leads to more satisfactory results. 

Finally, we define the average received power $P_r$ over $N$ hops as:
\begin{align}
    P_r = \frac{1}{N} \sum_{j=1}^N \|\bm{z}_j\|_2^2 = \frac{1}{N} \sum_{j=1}^N \|\mathbf{W}_j\bm{s}_j\|_2^2.
    \label{eq:P_avg_r}
\end{align}
Note that we are also interested in the average transmit power over the entire network, which is defined as:
\begin{equation}
    P_t = \frac{1}{N} \sum_{i=1}^{N} \|\mathbf{F}_i \bm{s}_i\|_2^2.
    \label{eq:P_avg_t}
\end{equation}
However, since we do not have access to $\mathbf{F}_i\bm{s}_i$ while training PrototypeNet, we use $P_r$ to train it as in \eqref{eq:optim_z}.

\begin{algorithm}[t!]
\caption{Optimization for the $i$-th precoding matrix $\mathbf{F}_i$.}\label{alg:calculate_Fi}
\begin{algorithmic}[1]
\State{{Input} $\{\mathbf{W}_i, \mathbf{H}_i, \tilde{\mathbf{S}}_i, \lambda_1, \eta, \epsilon\}$; {Output} $\mathbf{F}_i$}
\State{{Initialize} $\mathbf{F}_i = \mathbf{H}_i^{-1} \mathbf{W}_i,$}
\State{{Objective} $G_{\rm p} = \|\mathbf{W}_i - \mathbf{H}_i \mathbf{F}_i\|^2_F + \lambda_1 {\rm tr}(\tilde{\mathbf{S}}_i\mathbf{F}_i^\dagger\mathbf{F}_i)$}
\While{\textbf{True}}:

\State{{Calculate the gradient:} $\nabla \mathbf{F}_i = -2\mathbf{H}_i^\dagger \mathbf{W}_i + 2\mathbf{H}_i^\dagger \mathbf{H}_i \mathbf{F}_i + 2\lambda_1 \mathbf{F}_i \tilde{\mathbf{S}}_i$,}
\State{{Update:} $\mathbf{F}_i = \mathbf{F}_i - \eta \nabla \mathbf{F}_i$,}
\State{{New objective:} $G_{\rm c} = \|\mathbf{W}_i - \mathbf{H}_i \mathbf{F}_i\|^2_F + \lambda_1 {\rm tr}(\tilde{\mathbf{S}}_i\mathbf{F}_i^\dagger\mathbf{F}_i)$,}
\If{$|G_{\rm c} - G_{\rm p}| < \epsilon$}
\State{{Return} $\mathbf{F}_i$.}
\Else
\State{$G_{\rm p} = G_{\rm c}$.}
\EndIf

\EndWhile
\end{algorithmic}
\end{algorithm}

\subsection{Calculation of $\mathbf{F}_i$} \label{sec:IIIB}
Once the weights of PrototypeNet are obtained, at each realization of the network state, our goal will be to determine the precoding matrices to approximate the behavior of PrototypeNet over-the-air. Take the first hop as an example: since $\mathbf{H}_1$ is available at the source node, the precoding matrix $\mathbf{F}_1$ will be determined using $\mathbf{W}_1$ and $\mathbf{H}_1$. 

Note that each element of $\mathbf{H}_1$ follows an i.i.d. complex Gaussian distribution. It can be shown that $\mathbf{H}_1$ is full-rank with probability  one. Thus, the inverse matrix of $\mathbf{H}_1$ always exists. A simple least square criterion can be applied to obtain 
\begin{equation}
    \mathbf{F}_1 = \mathbf{H}^{-1}_1 \mathbf{W}_1.
    \label{eq:F_naive}
\end{equation}
In this case, since we have $\mathbf{H}_1\mathbf{F}_1 = \mathbf{W}_1$, the corresponding classification accuracy is identical to that of PrototypeNet. %However, the naive approach leads to overwhelmingly high transmit power $\|\mathbf{F}_1 \bm{s}_1\|^2_2$. This can be understood by the fact that the randomly generated $\mathbf{H}_1$ may be ill-conditioned and some singular values of $\mathbf{H}^{-\dagger}_1$ are extremely large. 
However, this naive approach fails to take the transmit power consumption $P_t$, defined in \eqref{eq:P_avg_t}, into account. In particular,  a slight deviation from $\mathbf{F}_1$ in \eqref{eq:F_naive} can reduce the power consumption significantly without sacrificing the classification accuracy.
To this end, we propose an alternative approach to determine $\mathbf{F}_1$ which is detailed as follows.

We formulate an optimization problem with an objective to achieve high classification accuracy while consuming low transmit power. For the former objective, since $\mathbf{W}_1$ has already been optimized, we expect the term $\|\mathbf{H}_1\mathbf{F}_1 - \mathbf{W}_1\|^2_F$ to be as small as possible. For the second one, we are interested in minimizing the expectation of the transmit power, i.e., $\mathbb{E}(\|\mathbf{F}_1 \bm{s}_1\|^2_2)$ over all $\bm{s}_1$. Note that $\mathbb{E}(\|\mathbf{F}_1 \bm{s}_1\|^2_2)$ can be alternatively expressed as:
\begin{equation}
    \mathbb{E}(\|\mathbf{F}_1 \bm{s}_1\|^2_2) = \mathbb{E}({\rm tr}(\bm{s}_1\bm{s}_1^\dagger \mathbf{F}_1^\dagger\mathbf{F}_1)) = {\rm tr}(\mathbf{S}_1\mathbf{F}_1^\dagger\mathbf{F}_1),
    \label{eq:power_expect}
\end{equation}
where $\mathbf{S}_1 \triangleq \mathbb{E}(\bm{s}_1\bm{s}_1^\dagger)$ and ${\rm tr}(\cdot)$ calculates the trace of a matrix. Given that the underlying image distribution, $P_{\mathbf{S}}$ is unknown, it is impossible to calculate the exact $\mathbf{S}_1$ term. Thus, we estimate $\mathbf{S}_1$ by averaging over the training set:
\begin{equation}
    \tilde{\mathbf{S}}_1 = \frac{1}{N_{tr}}\sum_{i=1}^{N_{tr}} \bm{s}_{1,i}\bm{s}_{1,i}^\dagger,
    \label{eq:covariance_est}
\end{equation}
where $N_{tr}$ denotes the number of images in the training set and $\bm{s}_{1, i}$ is the transformed input in \eqref{eq:s1} corresponding to the $i$-th image.

Finally, the optimization problem can be expressed as:
\begin{align}
    \min_{\mathbf{F}_1}  \underbrace{\|\mathbf{W}_1  - \mathbf{H}_1 \mathbf{F}_1\|^2_F + \lambda_1 {\rm tr}(\tilde{\mathbf{S}}_1\mathbf{F}_1^\dagger\mathbf{F}_1)}_{G},
    \label{eq:optim_formulate}
\end{align}
where $\lambda_1$ balances the classification accuracy and the transmit power consumption. We note that the naive solution $\mathbf{F}_1 = \mathbf{H}^{-1}_1 \mathbf{W}_1$ can be treated as a special case of \eqref{eq:optim_formulate} with $\lambda_1 = 0$.
It is easy to show that the optimization problem is convex, and can be solved via gradient descent. Specifically, for given $\mathbf{H}_1$ and $\mathbf{W}_1$,  $\mathbf{F}_1$ is updated  as follows:
\begin{align}
    \nabla \mathbf{F}_1 &= -2\mathbf{H}_1^\dagger \mathbf{W}_1 + 2\mathbf{H}_1^\dagger \mathbf{H}_1 \mathbf{F}_1 + 2\lambda_1 \mathbf{F}_1 \tilde{\mathbf{S}}_1, \notag \\
    \mathbf{F}_1 &= \mathbf{F}_1 - \eta \nabla \mathbf{F}_1,
    \label{eq:gd_F}
\end{align}
where $\nabla \mathbf{F}_1$ represents the gradient of $\mathbf{F}_1$ and $\eta$  is the step size. 
The optimization process can be terminated when the reduction in the objective function, $G$, defined in \eqref{eq:optim_formulate}  is smaller than $\epsilon$ in two consecutive steps. The overall optimization process is summarized in Algorithm \ref{alg:calculate_Fi}. We calculate the precoding matrix, $\mathbf{F}_i$ for each hop according to different statistics, $\tilde{\mathbf{S}}_i$ and channel realizations, $\mathbf{H}_i$.  The performance of the optimization algorithm is evaluated in terms of the classification accuracy and the average transmit power $P_t$  defined in \eqref{eq:P_avg_t}.

\subsection{Multiple-Block Transmission Strategy}
In a practical scenario, the relay nodes will have a limited number of antennas. This implies that the dimension of $\bm{s}_i$ may be prohibitively small resulting in degraded classification performance with a single transmission. In this subsection,  we will consider the transmission of $J$ channel inputs per input image so that the system with $M$ antennas will be capable of supporting $L$-dimensional inputs, $\bm{s}_i$. Without loss of generality, we assume $L = JM$.

To describe the system operation, we again take the source node as an example. After generating $\bm{s}_1$, the source node partitions it into $J$ blocks, denoted as $\bm{s}_1 = [\bm{s}_1^1, \ldots, \bm{s}_1^J]$. For the $j$-th block, the signal $\bm{s}_1^j$ is precoded by $\mathbf{F}_1^j$. We assume the channel matrix, $\mathbf{H}_1$ remains constant over the block of $J$ transmissions, and the received signal at the first relay can be expressed as $\bm{y}_1^j = \mathbf{H}_1 \bm{s}_1^j + \bm{n}_1^j$. By stacking the received signal for $J$ time slots, we obtain:
\begin{equation}
    \bm{y}_1 = (\mathbf{I}_J \otimes \mathbf{H}_1) {\rm diag}(\mathbf{F}_1^1, \ldots, \mathbf{F}_1^J) \bm{s}_1,
    \label{eq:y1_T}
\end{equation}
where $\otimes$ stands for the Kronecker product and ${\rm diag}(\cdot)$ denotes a diagonal matrix with diagonal elements being the matrices $\{\mathbf{F}_1^1, \ldots, \mathbf{F}_1^J\}$. 

To obtain the precoding matrices, we follow the same procedure discussed in Section \ref{sec:IIIA} and \ref{sec:IIIB}. In the training phase, for each hop, $J$ neural network weights, $\mathbf{W}_1^j \in \mathbb{C}^{M \times M}$ are obtained via end-to-end training. Then, given channel realization $\mathbf{H}_1$, we calculate $\mathbf{F}_1^j$  similarly to  \eqref{eq:optim_formulate}:
\begin{equation}
    \min_{\mathbf{F}_1^j} \|\mathbf{W}_1^j - \mathbf{H}_1 \mathbf{F}_1^j\|^2_F + \lambda_1 {\rm tr}(\tilde{\mathbf{S}}_1^j(\mathbf{F}_1^j)^\dagger\mathbf{F}_1^j),
    \label{eq:optim_formulate_t}
\end{equation}
where $\tilde{\mathbf{S}}_1^j$ denotes the expectation of $\bm{s}_1^j (\bm{s}_1^j)^\dagger$ as in \eqref{eq:covariance_est}.

We note that 
%the proposed multiple transmission strategy can be considered as dividing the input image into $J$ parts, and applying a separate neural network to each part. The outputs of these neural networks are combined and processed jointly by the FC layer at the destination.
in the considered scenario with $M$ antennas, there are $JM^2$ learnable parameters for each hop, which is smaller than that with $JM$ antennas where $(JM)^2$ parameters would be used. 
% As will be shown in the simulation part, the framework with $L$ antennas is outperformed by that with $M$ antennas.

\section{Numerical Experiments}\label{sec:IV}
\begin{figure*}
     \centering
     \begin{subfigure}{{0.64\columnwidth}}
         \centering
         \includegraphics[width=\columnwidth]{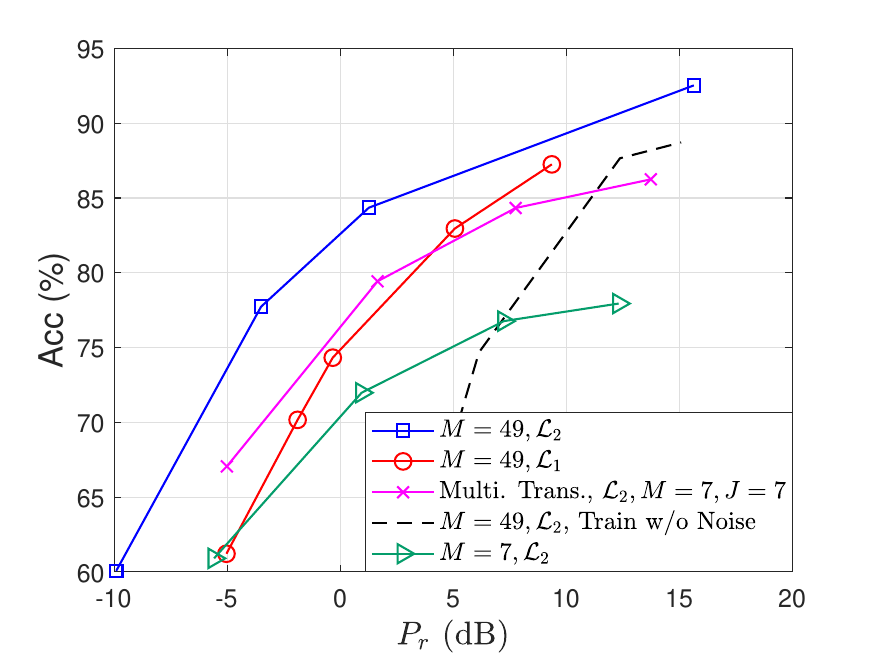}
         \caption{}
     \end{subfigure}
     \begin{subfigure}{{0.64\columnwidth}}
         \centering
         \includegraphics[width=\columnwidth]{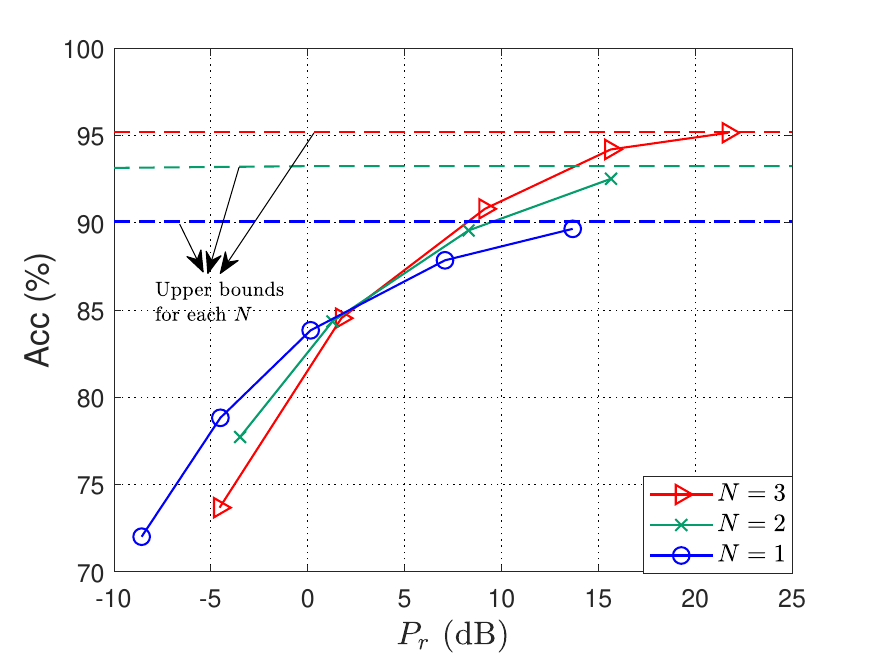}
         \caption{}
     \end{subfigure}
    %  \vspace{0.1cm}
     \begin{subfigure}{0.64\columnwidth}
         \centering
         \includegraphics[width=\columnwidth]{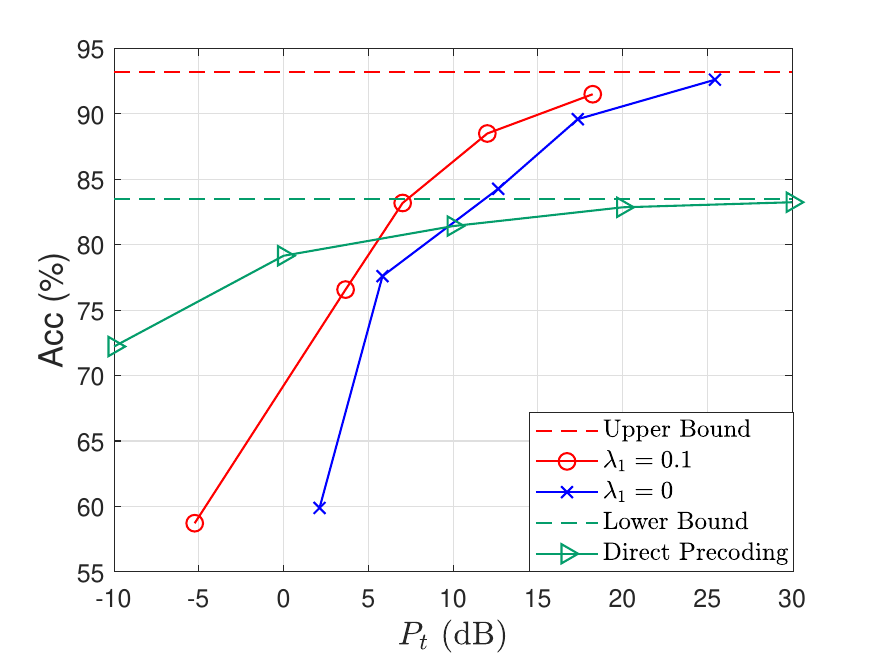}
         \caption{}
     \end{subfigure}
  \caption{Classification performance of the proposed scheme in terms of the average power consumption. (a) The performance obtained by the PrototypeNet with different loss functions, number of antennas, ($M$) and number of transmissions, ($J$). (b) The performance of the PrototypeNet with different number of hops ($N$). (c) Performance of multi-hop MIMO network compared with a direct precoding baseline.}
\label{fig:final_simu}
\end{figure*}

In this section, we evaluate the classification accuracy and the power consumption $P_t$ and $P_r$ of the proposed multi-hop MIMO network.
The images are chosen from the Fashion MNIST dataset where each of the $28\times 28$ greyscale images belongs to $K = 10$ classes. Unless specified otherwise, we consider the number of hops $N=2$, the number of antennas $M = 49$ and a noise variance of $\sigma^2 = 0.1$.  
To train PrototypeNet, we adopt the Adam optimizer with a learning rate of $5\times 10^{-4}$. 
The number of epochs and the batch size are set to $200$ and $32$, respectively.

\subsection{Performance Evaluation}
In Fig. \ref{fig:final_simu}(a), we study $P_r$ versus classification accuracy for the two different loss functions presented in \eqref{eq:optim_W} and \eqref{eq:optim_z}. PrototypeNet corresponding to different loss functions are trained with different $\lambda$ values. For each scenario, different $\lambda$ values correspond to different points on the accuracy-average power ($P_r$) trade-off curve.  It can be seen that the network trained using the second loss function, i.e., $\mathcal{L}_2$, strictly outperforms that using $\mathcal{L}_1$, which indicates that it is more effective to directly minimize the target power consumption. 
Thus, in the rest of the paper, all the results are based on $\mathcal{L}_2$ as the loss function. We also evaluate an alternative solution where PrototypeNet is trained without noise, yet tested in the presence of noise. The corresponding performance is shown in the figure with a dashed line. A significant performance degradation is observed, which shows the necessity of taking the noise into account during training. 
The multiple-block transmission strategy presented in Section \ref{sec:III} is also evaluated. In particular, we consider a relay network with $M=7$ antennas at each node. Naturally, this network leads to significant performance loss compared to $M = 49$ antennas as this would result in a neural network with a much smaller number of neurons at each layer. On the other hand, when  $J = 7$ transmissions per input image is considered over the network with $M = 7$ antennas, the classification accuracy increases significantly. However, it is still inferior to the case of $M = 49$. This is because the $M = J = 7$ setup has less number of learnable parameters and the equivalent neural network architecture is more restricted. 
%We also consider evaluating the performance of the wireless network where the precoder matrices, $\bm{F}_i$ are obtained by solving the optimization problem in \eqref{eq:optim_formulate}. In particular, we consider $\lambda_1 = \{0, 0.1\}$ which correspond to the dashed lines in the same figure. As can be seen, when $\lambda_1 = 0$, $\bm{H}_i\bm{F}_i$ is exactly the neural network weight $\bm{W}_i$, thus it coincides with the curve of the PrototypeNet. The dashed line corresponding to $\lambda_1 = 0.1$ is inferior to that of the PrototypeNet as its aims to minimize the transmit power.

In Fig. \ref{fig:final_simu}(b), we study 
the impact of the number of hops $N$  on the system performance.  Different PrototypeNet architectures corresponding to $N = 1, 2, 3$ are trained with different $\lambda$ values.  The dashed lines represent the accuracy upper bound achieved by the PrototypeNet architectures without noise. As can be seen, when $N$ is small, the accuracy upper bound is lower. This aligns with the intuition that, for a small $N$, we have less number of FC layers and the performance is limited due to the shallow neural networks. When the number of hops increases, the accuracy upper bound improves. 
We also observe that under a low $P_r$, the PrototypeNet with a larger $N$ is inferior to that with a small $N$. This is because when $N$ becomes larger, the noise accumulates over the hops leading to poorer results. However, as it can be seen when $P_r$ increases, the accuracy of the PrototypeNet approaches its upper bound.

In Fig. \ref{fig:final_simu}(c), we outline the effectiveness of the proposed scheme over multi-hop MIMO network by comparing it with a direct precoding baseline, detailed as follows. For the $i$-th hop, the channel realization $\mathbf{H}_i$ can be expressed using singular value decomposition (SVD): $\mathbf{H}_i = \mathbf{U}_i \mathbf{\Sigma}_i \mathbf{V}_i^\dagger$.
The transmitter ($\mathrm{R}_{i-1}$) precodes $\bm{s}_i$ using $\mathbf{V}_i$ and transmits it over the MIMO channel.  The $i$-th relay $\mathrm{R}_{i}$ combines the received signal $\bm{y}_i$ using $\mathbf{U}_i^\dagger$ and applies MMSE equalization to   generate an estimate of $\bm{s}_i$:
\begin{equation}
    \tilde{\bm{s}}_i = (\mathbf{\Sigma}_i^\dagger \mathbf{\Sigma}_i + \sigma^2 \mathbf{I}_M)^{-1}\mathbf{\Sigma}_i^\dagger \mathbf{U}_i^\dagger \bm{y}_i.
\end{equation}
The $\tilde{\bm{s}}_i$ is then power normalized and precoded to be transmitted over the $(i+1)$-th hop. Finally, the destination node estimates $\tilde{\bm{s}}_N$ which is fed to $f_c(\cdot)$ followed by the softmax function to generate the classification output.  We can observe in Fig. \ref{fig:final_simu}(c) that when $P_t$ increases, the proposed scheme outperforms the direct precoding baseline. It is also observed that superior performance is obtained with $\lambda_1 = 0.1$. This is because having a non-zero $\lambda_1$ value tries to minimize the transmit power as analyzed in Section \ref{sec:III}. The two dashed lines in the figure represent the results obtained via noiseless PrototypeNet with $N = 0$ and $N = 2$ FC layers, and they serve as the upper bounds of the direct precoding and the proposed scheme, respectively.
Note that the proposed scheme with $\lambda_1 = 0.1$ outperforms the baseline when $P_t > 5$ dB,  which verifies the effectiveness of PrototypeNet and the optimization method in Algorithm \ref{alg:calculate_Fi}.

\section{Conclusion}
In this paper, we study a novel OAC framework over a multi-hop MIMO network for image classification. For each hop, the relay processes its received signal and calculates the precoding matrix to imitate the FC layer obtained from a PrototypeNet which is trained in an end-to-end fashion.
Then, the precoding matrix for each hop is obtained by solving an optimization problem.
We further consider a practical scenario with a limited number of antennas and propose a multiple-block transmission strategy to significantly improve the classification performance.  
Numerical experiments verify the effectiveness of the proposed OAC scheme in terms of classification accuracy over a multi-hop MIMO network and show the superiority of the multiple-block transmission strategy.

\appendices

% \section{Experimental setup}\label{sec:AppA}
% \input{AppendixA.tex}

\bibliographystyle{IEEEtran}
\bibliography{References}

\begin{thebibliography}{1}
\providecommand{\url}[1]{#1}
\csname url@samestyle\endcsname
\providecommand{\newblock}{\relax}
\providecommand{\bibinfo}[2]{#2}
\providecommand{\BIBentrySTDinterwordspacing}{\spaceskip=0pt\relax}
\providecommand{\BIBentryALTinterwordstretchfactor}{4}
\providecommand{\BIBentryALTinterwordspacing}{\spaceskip=\fontdimen2\font plus
\BIBentryALTinterwordstretchfactor\fontdimen3\font minus \fontdimen4\font\relax}
\providecommand{\BIBforeignlanguage}[2]{{%
\expandafter\ifx\csname l@#1\endcsname\relax
\typeout{** WARNING: IEEEtran.bst: No hyphenation pattern has been}%
\typeout{** loaded for the language `#1'. Using the pattern for}%
\typeout{** the default language instead.}%
\else
\language=\csname l@#1\endcsname
\fi
#2}}
\providecommand{\BIBdecl}{\relax}
\BIBdecl

\bibitem{edge_comp_over}
J.~Chen and X.~Ran, ``Deep learning with edge computing: A review,'' \emph{Proc. IEEE}, vol. 107, no.~8, pp. 1655--1674, Aug. 2019.

\bibitem{image_retrieval}
M.~Jankowski, D.~Gündüz, and K.~Mikolajczyk, ``Wireless image retrieval at the edge,'' \emph{IEEE J. Sel. Areas Commun.}, vol.~39, no.~1, pp. 89--100, Jan. 2021.

\bibitem{oac_fed}
K.~Yang, T.~Jiang, Y.~Shi, and Z.~Ding, ``Federated learning via over-the-air computation,'' \emph{IEEE Trans. Wireless Commun.}, vol.~19, no.~3, pp. 2022--2035, Mar. 2020.

\bibitem{oac_amiri}
M.~Mohammadi~Amiri and D.~Gündüz, ``Machine learning at the wireless edge: Distributed stochastic gradient descent over-the-air,'' \emph{IEEE Trans. Signal Process.}, vol.~68, pp. 2155--2169, Mar. 2020.

\bibitem{oac_iot}
G.~Zhu, J.~Xu, K.~Huang, and S.~Cui, ``Over-the-air computing for wireless data aggregation in massive \textsc{IoT},'' \emph{IEEE Wireless Commun.}, vol.~28, no.~4, pp. 57--65, 2021.

\bibitem{oac_mimo}
G.~Zhu and K.~Huang, ``\textsc{MIMO} over-the-air computation for high-mobility multimodal sensing,'' \emph{IEEE IoT J}, vol.~6, no.~4, pp. 6089--6103, 2019.

\bibitem{air_ris}
G.~Sanchez \emph{et~al.}, ``\textsc{AirNN}: Over-the-air computation for neural networks via reconfigurable intelligent surfaces,'' \emph{IEEE/ACM Trans. Netw.}, vol.~31, no.~6, pp. 2470--2482, Dec. 2023.

\bibitem{air_fc}
G.~Reus-Muns, K.~Alemdar, S.~G. Sanchez, D.~Roy, and K.~R. Chowdhury, ``\textsc{AirFC}: Designing fully connected layers for neural networks with wireless signals,'' in \emph{MobiHoc '23}, New York, NY, USA, Oct. 2023, p. 71–80.

\bibitem{air_mimo}
Y.~Yang, Z.~Zhang, Y.~Tian, Z.~Yang, C.~Huang, C.~Zhong, and K.-K. Wong, ``Over-the-air split machine learning in wireless \textsc{MIMO} networks,'' \emph{IEEE J. Sel. Areas Commun.}, vol.~41, no.~4, pp. 1007--1022, Apr. 2023.

\end{thebibliography}

\end{document}